\documentclass[journal]{IEEEtran}

\usepackage{amsfonts}
\usepackage{bm}
\usepackage{amsmath}
\usepackage{amssymb}
\usepackage{enumerate}
\usepackage{color,xcolor}
\usepackage{graphicx}
\usepackage{multirow,tabularx}
\usepackage{subfigure}
\usepackage{hyperref}
\usepackage[compress]{cite}
\usepackage{algorithm}
\usepackage{algpseudocode}
\usepackage{subfigure}
\usepackage{graphicx}
\usepackage{caption2}
\usepackage{mathtools}

\ifCLASSINFOpdf
\else
\fi
\begin{document}

\title{Random NOMA With Cross-Slot Successive Interference Cancellation Packet Recovery}

\author{Zhaoji~Zhang,~Ying~Li,~\IEEEmembership{Member,~IEEE,}~Guanghui Song,~\IEEEmembership{Member,~IEEE,} Chau Yuen,~\IEEEmembership{Senior Member,~IEEE,} and Yong Liang Guan,~\IEEEmembership{Senior Member,~IEEE}
	\thanks{This work was supported in part by National Natural Science Foundation of China (NSFC) under Grant 61971333, NSFC under Grant 61671345, and A*STAR AME IAF-PP under Grant A19D6a0053.  (\emph{Corresponding Author: Ying Li.})}
	\thanks{Z. Zhang and Y. Li are with Xidian University, Xi’an, 710071, China
		(e-mail: zjzhang\_1@stu.xidian.edu.cn; yli@mail.xidian.edu.cn).}
	\thanks{G. Song and C. Yuen are with Singapore University of Technology and Design, Singapore 119613 (e-mail: ghsong2008@hotmail.com; yuenchau@sutd.edu.sg).} 
	\thanks{Y. L. Guan is with the School of Electrical and Electronic Engineering, Nanyang Technological University, Singapore 639798 (e-mail:, eylguan@ntu.edu.sg).}}
\maketitle
\begin{abstract}
Conventional power-domain non-orthogonal multiple access (NOMA) relies on precise power control, which requires real-time channel state information at transmitters. This requirement severely limits its application to future wireless communication systems. To address this problem, we consider NOMA without power allocation, where we exploit the random channel fading and opportunistically perform successive interference cancellation (SIC) detection. To mitigate the multi-user interference, we propose a random NOMA where users randomly transmit their data packets with a certain probability. Then a cross-slot SIC packet recovery scheme is proposed to recover transmitted data packets. We model the cross-slot SIC packet recovery as a Markov process, and provide a throughput analysis, based on which the sum rate is maximized by jointly optimizing the transmission probability and the encoding rate of users.
\end{abstract}
\begin{IEEEkeywords}
Random NOMA, cross-slot SIC, Markov process, sum rate.
\end{IEEEkeywords}
\IEEEpeerreviewmaketitle
\section{Introduction}
\IEEEPARstart{A}{s} a promising supporting technique for future wireless communication systems, the power-domain non-orthogonal multiple access (NOMA) enables simultaneous transmission of multiple data symbols over the same time-frequency resource, and therefore significantly improves the resource utilization efficiency.

In downlink NOMA \cite{NOMA1, NOMA2}, data symbols targeted for different devices are superimposed at the base station (BS) with different power, and then broadcasted to the devices. At each receiving device, different data symbols experience the same fading, so that successive interference cancellation (SIC) detection can be conducted to recover each data symbol. However, for uplink NOMA, symbols from different transmitting devices experience different random fading, which may result in SIC detection failure. In order to address this problem, uplink power control has been proposed \cite{NOMA3, NOMA4}, where the transmission power is allocated according to different path losses of different devices. In addition, the channel inversion technique \cite{NOMA5, NOMA6, NOMA7,NOMA8} exploits instantaneous channel state information at transmitters (CSIT) to perform transmission power allocation, so that each transmitted symbol can be received at designated power. Although power allocation is an important factor to improve the system sum rate, power allocation-based NOMA requires real-time CSIT and precise power control, which severely limits its application to the massive number of low-cost devices in future wireless communication \cite{servicetype}.

In this paper, we consider uplink NOMA without power allocation, where the SIC detection is performed opportunistically according to the random channel feature. To avoid severe packet collisions, we propose a random NOMA scheme, where users randomly transmit their data packets with a certain probability. Then the BS performs cross-slot SIC packet recovery to improve the recovery probability of previously failed data packets, so that higher system throughput can be achieved. Furthermore, we model the cross-slot SIC packet recovery as a Markov process, and the state-transition in this Markov process is analyzed to provide a throughput analysis. Based on this throughput analysis, we maximize the sum rate of the proposed random NOMA scheme by jointly optimizing the transmission probability and the encoding rate of users.

Note that the idea of random transmission in NOMA is originally from the slotted Aloha schemes \cite{CRDSA,IRSA,CSA,song,frameless,Capture,FirstCapture,LYT}. However, works in \cite{CRDSA,IRSA,CSA,song,frameless} do not consider SIC detection at the receiver, and data packets can only be recovered from collision-free slots. Therefore, \cite{CRDSA,IRSA,CSA,song,frameless} in fact belong to orthogonal multiple access (OMA). Compared with \cite{FirstCapture,LYT}, we propose a cross-slot SIC packet recovery to improve the recovery probability of transmitted data packets, as well as a throughput analysis based on a Markov process.
\section{System Model}\label{RAfading}
\begin{figure*}{
	\centering
	\includegraphics[width=1.45\columnwidth]{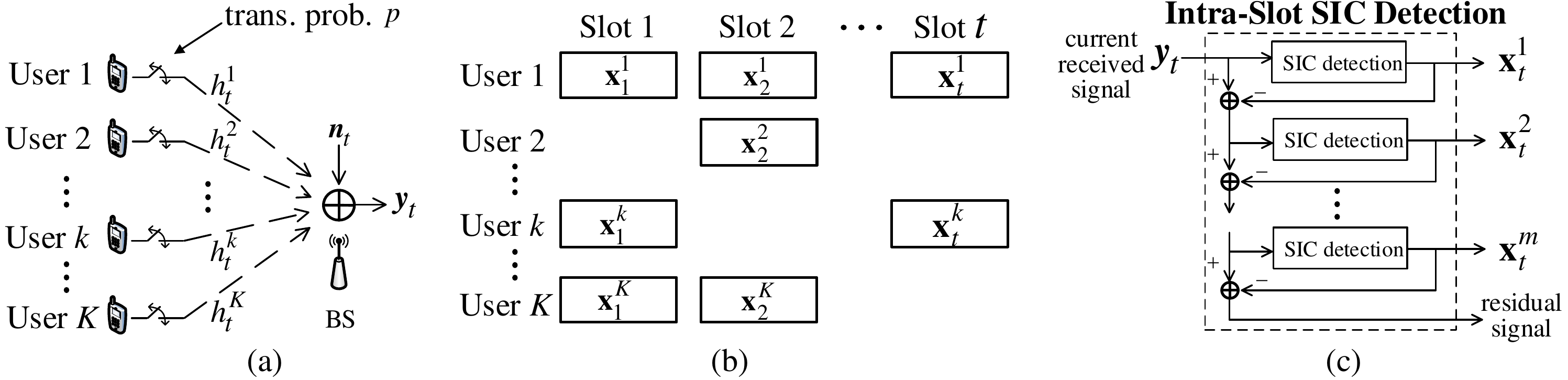}\vspace{-0.34cm}
	\caption{(a) $K$-user random NOMA system model.  (b) Random packet transmission in each slot. (c) Intra-slot SIC detection.}\vspace{-0.6cm}
	\label{model}}
\end{figure*}
\subsection{Random NOMA Transmission}
We illustrate our $K$-user random uplink NOMA system in Fig. \ref{model}(a). Each user has a list of data packets to be transmitted, and all the users attempt to transmit their data packets to a single BS over the same time-frequency resource. The entire transmission period is divided into time slots, where the duration of each time slot is configured to accommodate the transmission of one data packet.

Each transmitted data packet is assumed to experience a quasi-static Rayleigh fading, where the fading coefficient remains constant during one packet transmission, but varies independently from packet to packet. Different from conventional NOMA schemes \cite{NOMA3, NOMA4, NOMA5, NOMA6, NOMA7}, we assume that the channel state information (CSI) is only available to the BS, and no power allocation is adopted at the transmitter. Instead, we adopt equal transmission power $P_T$ for different users. In order to mitigate the multi-user interference, we propose the \emph{random} transmission scheme, i.e., in each time slot $t$, users choose to transmit data packets with a probability $p$, $0<p\leq1$. Then, the received signal $\mathbf{y}_t$ at time slot $t$ is
\begin{equation}\label{rcv}
\mathbf{y}_t=\sum_{k\in\mathcal{K}^+_t}\sqrt{P_T}\mathbf{x}_t^kh_t^k+\mathbf{n}_t
\end{equation}
where $\mathcal{K}^+_t$ is the set of active users in slot $t$, and $\mathbf{x}_t^k$ is the data packet transmitted from an active user $k\in\mathcal{K}^+_t$. In addition, each data symbol in $\mathbf{x}_t^k$ has unit power, and $h_t^k\sim\mathcal{CN}(0,1)$ represents the Rayleigh fading coefficient for user $k$ in slot $t$, and $\mathbf{n}_t$ is the additive white Gaussian noise, i.e., $\mathbf{n}_t \sim\mathcal{CN}(0,\sigma_n^2)$. In each time slot $t$, both $\mathcal{K}^+_t$ and $h_t^k, \forall k\in \mathcal{K}^+_t$ are assumed known at the BS. This active-user identification and CSI acquisition problem can be readily solved by inserting a unique pilot sequence into each user's data packet \cite{AMP}.
A schematic diagram of the random packet transmission is illustrated in Fig. \ref{model}(b). 

It should be emphasized that we consider NOMA without power allocation, and our random transmission strategy is employed for reducing the multi-user interference. However, this is not necessary for conventional power allocation-based NOMA, where SIC detection can be ideally conducted to eliminate the multi-user interference.
\subsection{Cross-Slot SIC Packet Recovery}
Different from the ideal SIC detection in power allocation-based NOMA, the SIC detection in our scheme is performed opportunistically due to the random fading. To improve the packet-recovery probability, the residual signal that contains unrecovered packets is stored in a collision buffer. Then the BS feeds back an acknowledge (ACK) message to the users whose packets are recovered, so that they can transmit new packets. No feedback is triggered by unrecovered packets, and corresponding users will randomly perform re-transmission until they detect the ACK message. Once a re-transmitted packet is recovered, its interference is cancelled from the collision buffer by cross-slot IC, so that the remaining packets in the buffer can be recovered. The details are explained below.

\underline{\textbf{Intra-slot SIC detection}}: Due to the random channel fading, data packets are received with different received power at the base station. Based on the different received power, SIC detection is performed opportunistically to recover as many packets as possible from the received signal $\mathbf{y}_t$. Without loss of generality, we assume $K^+$ active users in slot $t$, with $|h_t^1|^2>|h_t^{2}|^2>\ldots>|h_t^{K^+}|^2$. As shown in Fig. \ref{model}(c), the BS first recovers the packet $\mathbf{x}^1_t$ with the largest received power $P_T|h_t^1|^2$, and removes its interference $\sqrt{P_T}\mathbf{x}_t^1h_t^1$ from $\mathbf{y}_t$, while the signal of other packets is treated as noise. Then the BS successively recovers the packet  $\mathbf{x}^2_t$ with the second largest received power $P_T|h_t^2|^2$, and so on. In order to recover all the packets, the encoding rates of these $K^+$ users should satisfy the following $K^+$ inequalities simultaneously
\begin{equation}\label{SIC_condition}
R_k\leq\log_2\!\left(\!1\!+\!\frac{P_T|h^k_t|^2}{\sigma_n^2\!+\!\sum\limits_{i=k+1}^{K^+} \!P_T|h^i_t|^2}\right)\!\!, \ k=1,2,\ldots,K^+.
\end{equation}

Due to the random fading, (\ref{SIC_condition}) may not be satisfied for all $K^+$ users. Then the BS opportunistically finds a maximal number $m$ so that the first $m$ inequalities in (\ref{SIC_condition}) are satisfied. In this case, we call slot $t$ a $m$-recovery slot, since the BS can successively recover $m$ data packets from $\mathbf{y}_t$ by intra-slot SIC detection. For each of these $m$ recovered packets, an ACK message is fed back to the corresponding user, so that this user can start new transmission. On the other hand, no feedback is triggered by unrecovered packets, so that the active users that detect no ACK will perform re-transmission in future slots. In addition, the residual signal that contains $K^+-m$ unrecovered packets is called as a $(K^+-m)$-collision, and this residual signal is stored in a collision buffer for future cross-slot IC.

\underline{\textbf{Cross-slot IC}}: A recovered packet may be a new packet or a re-transmitted packet. If it is a re-transmitted packet, the BS will conduct cross-slot IC to cancel its interference from the buffer, since the re-transmitted copy contains exactly the same information as the previously-failed data packet. In this way, the interference is reduced for the remaining packets in the buffer, and a new round of intra-slot SIC detection is triggered to recover data packets from the buffer.

Overall, upon the reception of $\mathbf{y}_t$, the intra-slot SIC detection and cross-slot IC are performed iteratively to recover transmitted data packets. The proposed cross-slot SIC packet recovery is summarized in Algorithm \ref{alg:receiverSIC}.
\begin{algorithm}[t]
	\caption{$K$-User Cross-Slot SIC Packet Recovery}
	\label{alg:receiverSIC}
	\textbf{Initialize:} Let $t=$ the index of the current slot
	
	1:\textbf{if} slot $t$ is $m$-recovery with $m>0$ \textbf{then}
	
	2: \ \ \ \emph{Intra-slot SIC detection} in slot $t$.
	
	3: \ \ \ Store residual signal in the buffer.
	
	4: \ \ \ ACK feedback for each recovered packet.
	
	5: \ \ \ \emph{Cross-slot IC} (if retransmitted packets are recovered).
	
	6: \ \ \ \textbf{while} any buffer slot $t^\prime$ is $m^\prime$-recovery ($m^\prime>0$) \textbf{do}
	
	7: \ \ \ \ \ \ \emph{Intra-slot SIC detection} in slot $t^\prime$.
	
	8: \ \ \ \ \ \ ACK feedback for each recovered packet.
	
	9: \ \ \ \ \ \  \emph{Cross-slot IC} (if retransmitted packets are recovered).
	
	10: \ \ \textbf{end while}
	
	11:\textbf{else}
	
	12:\ \ \ Store the received signal into the collision buffer.
	
	13:\textbf{end if}
\end{algorithm}
\section{Throughput Analysis}\label{twouserana}
In this section, we model the cross-slot SIC packet recovery as a Markov process, and provide a throughput analysis. Due to the analysis complexity, we focus on the two-user NOMA with $K=2$, which is the most common case for NOMA \cite{NOMA1,NOMA2,NOMA3}. For analytical convenience, we assume the same encoding rate $R$ for all the users, i.e., $R_k=R$ for $k=1,2$.
\begin{figure}
	\centering
	\includegraphics[width=1\linewidth]{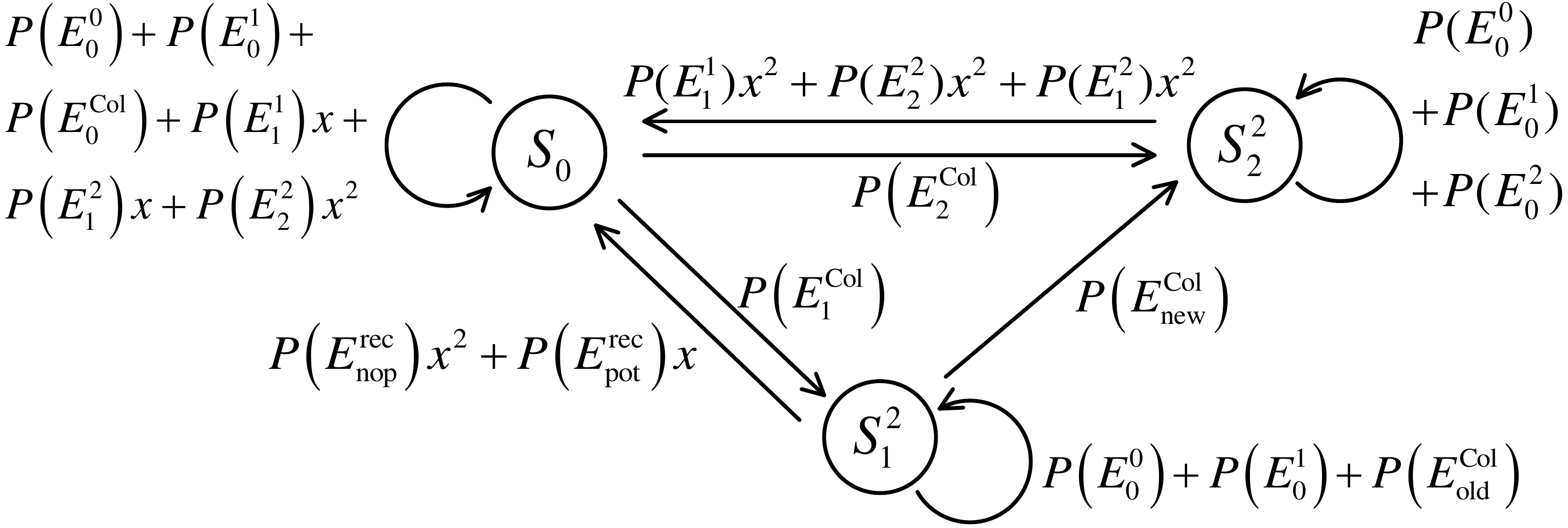}\vspace{-0.44cm}
	\caption{State transition diagram of the Markov process for the two-user case. When $x=1$, the edge label is equal to the state-transition probability of the Markov process.}
	\label{MarkovChain}\vspace{-0.5cm}
\end{figure}

In order to describe the state of the collision buffer, we first define the concept of \emph{potential} packets. In a 2-collision residual signal, an unrecovered packet $\mathbf{x}^k$ is termed a \emph{potential} packet if its fading coefficient satisfies
\begin{equation}\label{potential}
|h^k|^2>(2^R-1)\sigma_n^2/P_T.
\end{equation}

By definition, a potential packet $\mathbf{x}^k$ can be recovered from the collision buffer if the other packet in the same buffer slot is cancelled by cross-slot IC. Although the physical collision buffer may store different types of collisions, we only consider three effective buffer states $S_0$, $S^2_2$, and $S^2_1$, which are characterized by the number of different potential packets. $S_0$ is the state where the collision buffer contains no potential packet. On the other hand, the collision buffer contains two different potential packets in state $S^2_2$, while it is in state $S^2_1$ if there is only one potential packet in the collision buffer.

The cross-slot SIC packet recovery is modelled as a Markov process, since it only depends on the current buffer state and the received signal in current slot. The state-transition diagram of this Markov process is shown in Fig. \ref{MarkovChain}. Each state-transition is triggered by specific events occurring in the current slot, which are defined in Table. \ref{event}. Then the state-transition probability is derived by adding up the probabilities of the events resulting in this transition. For example, state $S^2_2$ will transit to state $S_0$ if any packet is recovered in the current slot. Therefore, this state-transition probability is equal to the sum of the probabilities of $E^1_1$, $E^2_2$, and $E^2_1$. The event probabilities are derived in Appendix \ref{prob}.

Each state-transition in Fig. \ref{MarkovChain} is labelled by a state-transition polynomial, where the power of $x$ represents the recovered packet number by corresponding events. Take the state-transition from $S^2_1$ to $S_0$ as an example. If $E^\text{rec}_\text{nop}$ occurs, the non-potential packet will be first recovered in the current slot. Then after cross-slot IC, the potential packet in the collision buffer can also be recovered. Therefore, a total number of two packets are recovered, and we have the term $P(E^\text{rec}_\text{nop})x^2$ in the state-transition polynomial. On the other hand, if $E^\text{rec}_\text{pot}$ occurs, the potential packet will be recovered in current slot, while the non-potential packet cannot be recovered even after cross-slot IC. Therefore, only one packet is recovered, and we have the term $P(E^\text{rec}_\text{pot})x$. Define a state-space vector $\mathbf{S}=[S_0\ S^2_2\ S^2_1]$, with $\mathbf{S}(i)$ as its $i$-th element. Then the state-transition polynomial matrix $\mathbf{P_{\text{st}}}(x)$ can be obtained in (\ref{stmatrix}) by replacing its $(i,j)$-th entry with the state-transition polynomial from state $\mathbf{S}(i)$ to state $\mathbf{S}(j)$. If $x$ is set to 1 in (\ref{stmatrix}), we will obtain the conventional state-transition matrix $\mathbf{P_{\text{st}}}=\mathbf{P_{\text{st}}}(x)|_{x=1}$ for the Markov process.

The transition throughput matrix $\mathbf{P_{\text{tt}}^{\prime}}$ is derived to indicate the expected number of recovered packets in each transition
\begin{table}
	\footnotesize
	\caption{Events occurring in the current slot.}
	\centering
	\begin{tabular}{|c|c|}
		\hline
		Event&Interpretation\\
		\hline
		$E^n_m$&$n$ active users with $m$-recovery\\
		\hline
		$E^\text{rec}_\text{nop}$&the non-potential packet in buffer is recovered\\
		\hline
		$E^\text{rec}_\text{pot}$&only the potential packet in buffer is recovered\\
		\hline
		$E^\text{Col}_j$&2-collision occurs with $j$ potential packets\\
		\hline
		$E^\text{Col}_\text{new}$&2-collision occurs with a new potential packet\\
		\hline
		$E^\text{Col}_\text{old}$&2-collision occurs without a new potential packet\\
		\hline
	\end{tabular}
	\label{event}\vspace{-0.58cm}
\end{table}
\begin{figure*}
	\begin{equation}\label{stmatrix}
	\mathbf{P_{\text{st}}}(x)=\left[
	{\begin{array}{*{20}{c}}
		P(E^0_0)+P(E^1_1)x+P(E^1_0)+P(E^2_2)x^2+P(E^2_1)x+P(E^\text{Col}_{0})&P(E^\text{Col}_{2})&P(E^\text{Col}_{1})\\
		P(E^1_1)x^2+P(E^2_2)x^2+P(E^2_1)x^2&P(E^0_0)+P(E^1_0)+P(E^2_0)&0\\
		P(E^\text{rec}_\text{nop})x^2+P(E^\text{rec}_\text{pot})x&P(E^0_0)+P(E^1_0)+P(E^\text{Col}_\text{old})&P(E^\text{Col}_\text{new})
		\end{array}}
	\right]\vspace{-0.1cm}
	\end{equation}\hrulefill\vspace{-0.5cm}
\end{figure*}
\begin{equation}\label{SlotRec}
\mathbf{P_{\text{tt}}^{\prime}}\!\overset{\Delta}{=}\!\!\!\left.\frac{\partial \mathbf{P_{\text{st}}}(x)}{\partial x}\right|_{x=1}\!\!\!=\!\!\!\left[
{\begin{array}{*{20}{c}}
	\!\!P(E^1_1)\!+\!P(E^2_1)\!+\!2P(E^2_2)&\!\!\!0\!\!\!\!&0\!\!\!\!\\
	\!\!2P(E^1_1)\!+\!2P(E^2_2)\!+\!2P(E^2_1)&\!\!\!0\!\!\!\!&0\!\!\!\!\\
	\!\!2P(E^\text{rec}_\text{nop})\!+\!P(E^\text{rec}_\text{pot})&\!\!\!0\!\!\!\!&0\!\!\!\!
	\end{array}}
\right]\!\!
\end{equation}
where $\mathbf{P_{\text{tt}}^{\prime}}(i,j)$ indicates the expected number of recovered packets in the transition from state $\mathbf{S}(i)$ to state $\mathbf{S}(j)$.

Denote $\mathbf{p}^{\infty}_{S}=[p^{\infty}_{0}\  p^{2\infty}_{2}\ p^{2\infty}_{1}]$ as the steady-state distribution vector, which indicates the steady-state probabilities of different states. The distribution vector $\mathbf{p}^{\infty}_{S}$ is obtained by solving the equation $\mathbf{p}^{\infty}_{S}\mathbf{P_{\text{st}}}=\mathbf{p}^{\infty}_{S}$ under the constraint $p^{\infty}_{0}+p^{2\infty}_{2}+p^{2\infty}_{1}=1$. Finally, the throughput $T$, which is the expected number of recovered packets in each slot as the total transmission slot approaches infinity, is derived as follows
\begin{equation}\label{throughput}
T=\mathbf{p}^{\infty}_{S}\mathbf{P_{\text{tt}}^{\prime}}\mathbf{c},\vspace{-0.2cm}
\end{equation}
where $\mathbf{c}=[1,1,1]^T$ sums up all the elements in $\mathbf{p}^{\infty}_{S}\mathbf{P_{\text{tt}}^{\prime}}$.

\section{Sum Rate Maximization}\label{capability}
The sum rate $R_s\overset{\Delta}{=}RT$ is defined as the expected amount of recovered information from each received symbol, and the maximization problem for $R_s$ is formulated as,
\begin{equation}\label{Opti}
\begin{split}
&\mathop {\text{max}}\limits_{p,R}: R_s=R\mathbf{p}^{\infty}_{S}\mathbf{P_{\text{tt}}^{\prime}}\mathbf{c}\\
&s.t.\ \ \ \  0< p\leq1, \ 0<R
\end{split}
\end{equation}
According to Section \ref{twouserana}, the throughput $T$ is a function of the encoding rate $R$ and the transmission probability $p$. Then the sum rate is maximized by jointly optimizing $R$ and $p$. 

The optimal solution ($p^*,R^*$) to (\ref{Opti}) is only determined by $K$ and noise level ${\sigma_n^2}$. Therefore, the BS does not need to feed back ($p^*,R^*$) for each transmission. In addition, we can optimize ($p,R$) offline for different $K$ and $\sigma_n^2$, and then store the optimization results at the BS. In this way, optimizing ($p,R$) will not cost any online computational complexity.
\subsection{Two-User Case}\label{twousersumrate}
\begin{figure}
	\centering
	\includegraphics[width=0.5\textwidth]{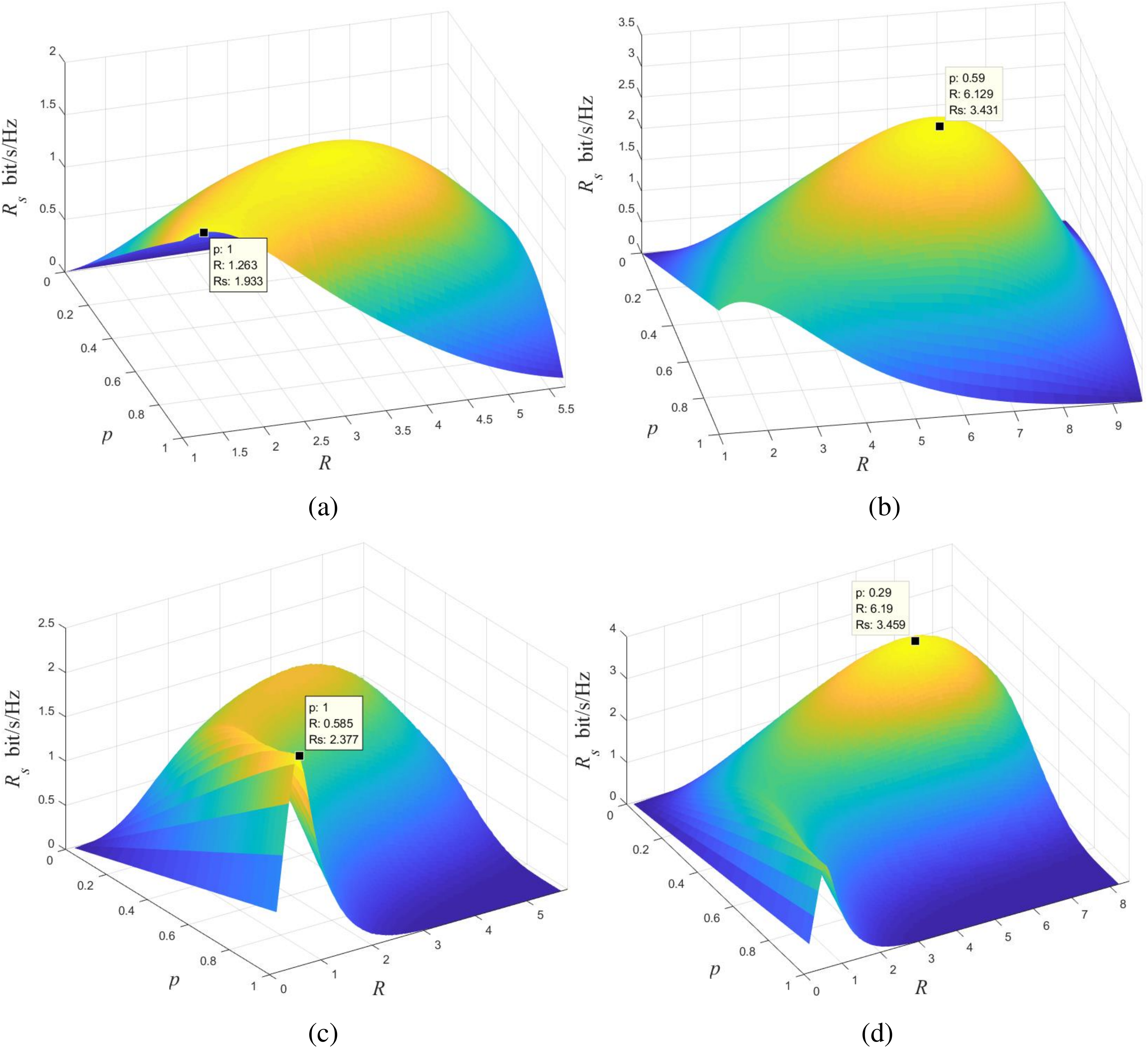}\vspace{-0.4cm}
     \caption{$R_s$ with different $p$ and $R$. For two-user case, we have (a) ($p^*=1.0, R^*=1.263, R_s^*=1.933$) under $B=15$dB (b) ($p^*=0.59, R^*=6.129, R_s^*=3.431$) under $B=25$dB. For $K=5$, we have (c) ($p^*=1.0, R^*=0.585, R_s^*=2.377$) under $B=15$dB (d) ($p^*=0.29, R^*=6.19, R_s^*=3.459$) under $B=25$dB. Simulation results averaged from $10^3$ experiments, with $n_s=200$ slots in each experiment.\vspace{-0.4cm}}\label{twouser}
\end{figure}
Since $R_s$ is a non-convex function of $R$ and $p$, we resort to a numerical solution to (\ref{Opti}). That is, according to the throughput analysis in Section \ref{twouserana}, the optimal solution $(p^*,R^*)$ is exhaustively searched for the maximized sum rate $R_s^*$.

Under $B=15$dB and $25$dB, the sum rate $R_s$ is illustrated as a function of $R$ and $p$ in Fig. \ref{twouser}(a)-(b). It is shown in Fig. \ref{twouser}(a) that when $B=15$dB, the maximized sum rate $R_s^*$ is equal to $1.933$, which is achieved by $(p^*=1,R^*=1.263)$. In this case, $R_s$ is maximized by deterministic transmission ($p^*=1$) with relatively low encoding rate. By contrast, when $B$ increases to $25$dB in Fig. \ref{twouser}(b), $R_s^*$ increases to 3.431, with $(p^*=0.59,R^*=6.129)$. In this case, $R_s$ is maximized by random transmission with higher encoding rate.
\subsection{$K$-User Case}
Theoretically, the throughput analysis and problem formulation (\ref{Opti}) can be extended to the general $K$-user case. However, the size of the state space increases dramatically for the Markov process with a large $K$, which greatly complicates the analysis. As an alternative, we simulate the cross-slot SIC packet recovery for $K$-user random NOMA, and find the optimal transmission probability and encoding rate ($p^*,R^*$) that maximize sum rate $R_s$. Here, $R_s$ is the average recovered information from each received symbol during $n_s$ slots.

Take $K=5$ as an example, and we illustrate $R_s$ under $B=15$dB and $25$dB in Fig. \ref{twouser}(c)-(d). Similarly, when $B=15$dB, the maximal sum rate $R^*_s=2.377$ is achieved by deterministic transmission ($p^*=1$) and low encoding rate ($R^*=0.585$) in Fig. \ref{twouser}(c). In contrast, when $B=25$dB, the maximal sum rate $R^*_s=3.459$ is achieved by random transmission ($p^*=0.29$) and higher encoding rate ($R^*=6.19$) in Fig. \ref{twouser}(d).
\begin{figure}
	\centering
	\includegraphics[width=0.49\textwidth]{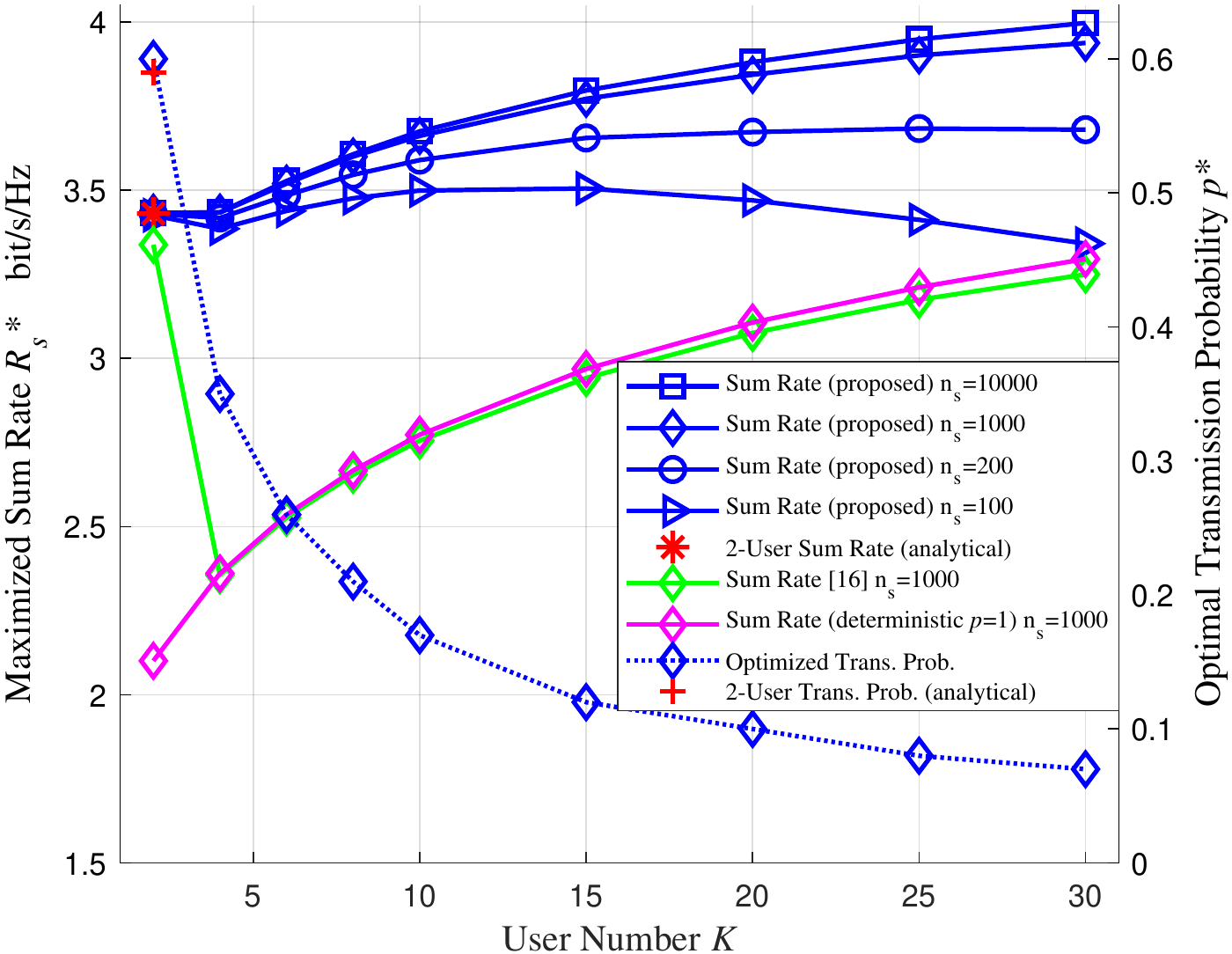}\vspace{-0.5cm}
	\caption{$R_s^*$ (solid lines) and $p^*$ (dashed line) for different $K$ under $B=25$dB. For our proposed random NOMA, $p^*$ decreases with increasing $K$, while $R^*\approx6.1$ (not plotted) for different $K$. Results are averaged from $10^3$ experiments.\vspace{-0.4cm}}\label{Kuser}
\end{figure}

The maximized sum rate $R^*_s$ for different $K$ is shown in Fig. \ref{Kuser} under $B=25$dB. For comparison, the performances of the random transmission scheme without cross-slot IC \cite{LYT} and the deterministic transmission scheme ($p=1$) are also plotted. The proposed random NOMA with its cross-slot SIC packet recovery outperforms the other two schemes. Especially for $K\geq4$, compared with the other schemes, our proposed random NOMA scheme could improve $R_s$ by $0.64\sim1.07$. In addition, as $K$ increases, the optimal transmission probability $p^*$ (dashed line) of our proposed scheme decreases. The optimal encoding rate $R^*$ (not plotted in Fig. \ref{Kuser}) approximately equals $6.1$ for different $K$. When $K=2$, the simulation results in Fig. \ref{Kuser} agree well with the analytical results in Fig. \ref{twouser}(b). When $K\geq10$, as $n_s$ increases from 100 to 10000, $R^*_s$ is improved and approaches the asymptotic performance with infinite $n_s$. Furthermore, we observe that with optimized solution ($p^*,R^*$), the buffered collision number is approximately equal to $K/2$ for different $K$. That is, only limited memory is required to store failed packets at the buffer.
\section{Conclusions}\label{conclusions}
We considered uplink NOMA without power allocation. A random NOMA scheme was proposed, where users randomly transmits data packets with certain probability to avoid severe collisions. Then the cross-slot SIC packet recovery was proposed to improve the packet-recovery probability. We modeled the cross-slot SIC packet recovery as a Markov process, and provided a throughput analysis. Based on this throughput analysis, the sum rate was maximized by jointly optimizing the transmission probability and the encoding rate of users.
\begin{appendices}
\section{Event Probabilities}\label{prob}
For notational simplicity, we denote the received SNR of $\mathbf{x}^k$ from user $k$ as $b_k\overset{\Delta}{=}P_T|h^k|^2/\sigma_n^2$. Then $b_k$ follows the exponential distribution with probability density function
\begin{equation}\nonumber
f(b_k|B)=\frac{1}{B}\exp{(-b/B)},\ \ b>0
\end{equation}
where $B=P_T/\sigma_n^2$ is the average received SNR. Furthermore, we denote $\rho_{th}\overset{\Delta}{=}2^{R}-1$, and assume $R\geq1$ for subsequent analysis. Then, the event probabilities are derived below.
	
\underline{$\mathbf{E^n_m}$}: The probability that $n$ out of $K$ users are active is equal to $C^K_np^n(1-p)^{K-n}$. According to (\ref{SIC_condition}), the current slot is $m$-recovery if the received SNR satisfies the following constraints
\begin{equation}\nonumber
\begin{split}
&\eta^n_k<b_k<\infty,\ \ k=1, 2,\ldots, m, \ \ b_{m+1}<b_{m+1}<\eta^n_{m+1}\\
&b_{j+1}<b_{j}<\infty,\ \ j=m+2,\ldots, n
\end{split}
\end{equation}
where $\eta^n_k\overset{\Delta}{=}\rho_{th}(1+\sum\limits_{i=k+1}^{n}b_i)$ is the SNR threshold to recover $\mathbf{x}^k$. Since there are totally $n!$ different permutations of $\{b_1,b_2,\ldots,b_n\}$, we have
\begin{equation}\label{generalint}
\begin{split}
&P(E^n_m)=C^K_np^n(1-p)^{K-n}\times n!\times\\
&\underbrace{\int\limits_{b_n\!>\!0}^{b_n\!<\!\infty}\!\!\!\!\cdots\!\!\!\!\int\limits^{b_{m\!+\!2}\!<\!\infty}_{b_{m\!+\!2}\!>\!b_{m\!+\!3}}\int\limits_{b_{m\!+\!1}\!>\!b_{m\!+\!2}}^{b_{m\!+\!1}\!<\!\eta^n_{m\!+\!1}}}_{n-m\ \text{unrecovered packets}}\underbrace{\int\limits_{b_{m}\!>\!\eta^n_{m}}^{b_{m}<\infty}\!\!\!\!\cdots\!\!\!\!\int\limits_{b_{2}\!>\!\eta^n_{2}}^{b_{2}<\infty}\int\limits_{b_{1}\!>\!\eta^n_{1}}^{b_{1}<\infty}}_{m\ \text{recovered packets}}\!\prod\limits_{i=1}^{n}\!\!f(b_i|B)db_1\!\cdot\!\cdot\!\cdot\! b_n.
\end{split}
\end{equation}
According to (\ref{generalint}), we have the following results
\begin{equation}\nonumber
\begin{split}
P(E^0_0)&=(1-p)^2,\ P(E^1_1)=2p(1-p)\exp\left(-\frac{\rho_{th}}{B}\right),\\
P(E^1_0)&=2p(1-p)\left[1-\exp\left(-\frac{\rho_{th}}{B}\right)\right],\\
P(E^2_2)&=\frac{2p^2}{1+\rho_{th}}\exp(-\frac{\rho_{th}(2+\rho_{th})}{B}),\\
P(E^2_1)&=\frac{2p^2}{1+\rho_{th}}\left[\exp(-\frac{\rho_{th}}{B})-\exp(-\frac{\rho_{th}(2+\rho_{th})}{B})\right],\\
P(E^2_0)&=p^2-\frac{2p^2}{1+\rho_{th}}\exp(-\frac{\rho_{th}}{B}).\\
\end{split}
\end{equation}
\underline{$\mathbf{E^\text{Col}_j}$}: Similar to (\ref{generalint}), we derive $P(E^\text{Col}_j)$ as follows,
\begin{equation}\nonumber
\begin{split}
P(E^\text{Col}_2)=&2!p^2\int_{\rho_{th}}^\infty\int_{b_2}^{\eta^2_1}f(b_1|B)f(b_2|B)db_1db_2\\
=&p^2\exp(-\frac{2\rho_{th}}{B}) \!-\! \frac{2p^2}{1+\rho_{th}}\exp(-\frac{\rho_{th}(2+\rho_{th})}{B}),\\
P(E^\text{Col}_1)=&2!p^2\int_0^{\rho_{th}}\int_{\rho_{th}}^{\eta^2_1}f(b_1|B)f(b_2|B)db_1db_2\\
=&\frac{2p^{2}\rho_{th}}{1+\rho_{th}}\exp(-\frac{\rho_{th}}{B})-2p^2\exp(-\frac{2\rho_{th}}{B})\\
&+\frac{2p^2}{1+\rho_{th}}\exp(-\frac{\rho_{th}(2+\rho_{th})}{B}),\\
P(E^\text{Col}_0)=&2!p^2\int_0^{\rho_{th}}\int^{\rho_{th}}_{b_2}f(b_1|B)f(b_2|B)db_1db_2\\
=&p^2[1-\exp(-\frac{\rho_{th}}{B})]^2.
\end{split}
\end{equation}
\underline{$\mathbf{E^\text{rec}_\text{nop}}$, $\mathbf{E^\text{rec}_\text{pot}}$, $\mathbf{E^\text{Col}_\text{new}}$, $\mathbf{E^\text{Col}_\text{old}}$}: These events can be divided into independent sub-events. For example, $E^\text{rec}_\text{pot}$ occurs if (a) $E^1_1$ or $E^2_1$ occurs and (b) recovered packet in current slot is the potential packet in buffer. If condition (a) is satisfied, condition (b) holds with probability $1/2$. Therefore, we have
\begin{equation}\nonumber
P(E^\text{rec}_\text{pot})={\left(P(E^1_1)+P(E^2_1)\right)}/{2}.
\end{equation}
Similarly, according to the definitions in Table. \ref{event}, we have
\begin{equation}\nonumber
\begin{split}
P(E^\text{rec}_\text{nop})&={\left(P(E^1_1)+P(E^2_1)\right)}/{2}+P(E^2_2),\\
P(E^\text{Col}_\text{new})&={P(E^\text{Col}_1)}/{2}+P(E^\text{Col}_2),\\ P(E^\text{Col}_\text{old})&={P(E^\text{Col}_1)}/{2}+P(E^\text{Col}_0).
\end{split}
\end{equation}
\end{appendices}

\ifCLASSOPTIONcaptionsoff
  \newpage
\fi



%

\end{document}